\documentclass[12pt]{iopart}

\usepackage{url}
\usepackage{graphicx}
\usepackage[square,numbers]{natbib}
\usepackage{setstack}
\usepackage{iopams}

\eqnobysec

\begin{document}

\title[Parameter identification with BioPARKIN]{Parameter identification in large kinetic networks with BioPARKIN}

\author{T Dierkes$^{1}$,
        S R{\"o}blitz$^{1}$, 
        M Wade$^{2}$, 
        P Deuflhard$^{1,3}$
       } 

\address{$^{1}$Zuse Institute Berlin,\\
         $^{\ }$ Takustra{\ss}e 7, 14195 Berlin-Dahlem, Germany}

\address{$^{2}$Heidelberg University Hospital, Radiology Clinic, Department for Nuclear Medicine,\\ 
        $^{\ }$Im Neuenheimer Feld 400, 69120 Heidelberg, Germany}

\address{$^{3}$Free University Berlin, Institute of Mathematics, Num.~Math.~/ Sci.~Comp.,\\ 
        $^{\ }$Arnimallee 6, 14195 Berlin-Dahlem, Germany}

\ead{susanna.roeblitz@zib.de}

\begin{abstract}
\paragraph{Motivation.}
Modelling, parameter identification, and simulation play an important role in
systems biology. Usually, the goal is to determine parameter values that minimise 
the difference between experimental measurement values and model predictions 
in a least-squares sense.  Large-scale biological networks, however, 
often suffer from missing data for parameter identification.  Thus, the 
least-squares problems are rank-deficient and solutions are not unique.  
Many common optimisation methods ignore this detail because they do not 
take into account the structure of the underlying inverse problem.  
These algorithms simply return a ``solution'' without additional information 
on identifiability or uniqueness.  This can yield misleading results, 
especially if parameters are co-regulated and data are noisy.

\paragraph{Results.}
The Gauss-Newton method presented in this paper monitors the numerical rank 
of the Jacobian and converges locally, for the class of adequate problems, 
to a solution that is unique within the subspace of identifiable parameters. 
This method has been implemented in BioPARKIN, a software package that 
combines state-of-the-art numerical algorithms with compliance to system 
biology standards, most importantly SBML, and an accessible interface.

\paragraph{Availability.}
The software package BioPARKIN is available for download at \\
\ \url{http://bioparkin.zib.de} .

\end{abstract}

\maketitle

\section{Introduction}

Following \citep{Schuppert10}, there are two main modelling approaches 
in systems biology.  On one hand, there exist detailed models for isolated 
parts of a system.  The states and model parameters of such systems are 
generally well-defined, but the system is far from being closed and there 
are great variations in the environmental conditions.  On the other hand, 
large-scale networks are more closed, but suffer from missing data for 
parameter identification.  Biological data, however, often indicate that 
parameters are correlated, and that a system's behaviour can be characterised 
by a few control parameters.  In contrast to parameter optimisation, 
{\em parameter identification} not only aims at the determination of parameter 
values from given measurement data, but also on the detection of dependencies 
between parameters.  As stated in \citep{Schuppert10}, the identification of all 
control parameters which allow a proper characterisation of the states 
of a biological system, is by no means trivial and, at least for most 
applications, an open problem. 

Modelling, parameter estimation and simulation of biological systems have 
become part of modern systems biology toolboxes. Unfortunately, many of 
these programs are based on inefficient or mathematically outdated algorithms. 
To counteract this problem, we have developed the software package 
BioPARKIN$^{1}$\footnote[0]{$^{1}$ {\bf Bio}logy-related {\bf par}amater identification in large {\bf kin}etic networks} 
\citep{Die11}. This software is a renewed version of the former codes 
LARKIN 
\citep{DdBa:81} 
and PARKIN 
\citep{Nowak85}, which have successfully been applied in chemical industry 
for more than 20 years \citep{Deuflhard1986}.

BioPARKIN combines a basis of long-standing mathematical principles with 
compliance to system biology standards, most importantly SBML \citep{cornish2003systems}, 
and an accessible interface. The SBML format is one of the most important standards in 
systems biology to facilitate collaboration of researchers at all levels (physicians, 
biologists, mathematicians, etc.).  The interface strives to wrap complicated structures 
and settings (especially with regard to the numerical back-end) into an user-friendly 
package that can be used correctly by non-mathematicians.

BioPARKIN is split into two parts -- the numerical library PARKINcpp and the graphical 
user interface (GUI) -- in order to achieve several advantages. The crucial, yet 
computation-intensivenumerical algorithms are embedded in an efficient C++ library 
while the GUI is coded in Python which enables rapid interface changes when adapting 
the user interface to new insights into user behaviour.  Another important advantage is 
the independent availability of the PARKINcpp library for use in other related projects. 
Both parts are available under the LPGL which is a flexible open-source license allowing 
for the use of the software in both open and closed (i.e.~commercial) projects.

The core of PARKINcpp and its unique feature is the solver 
NLSCON for {\bf n}onlinear {\bf l}east-{\bf s}quares with 
{\bf con}straints \citep{nlscon}.  This Gauss-Newton type 
method is especially suited for rank-deficient problems \citep{Deu04}. 
NLSCON requires, however, some user specified input such as 
threshold values for species and parameters, or a threshold 
value for rank decision.  In order to choose reasonable 
values and to obtain reliable results, it is indispensable 
to understand the foundations of the algorithm. 
This paper therefore aims at giving an overview of the 
functionality and implementation of NLSCON within BioPARKIN.

The article is organised as follows. We start with 
the problem definition in Section~\ref{sec:approach}.  
In Section~\ref{sec:methods} we explain our method to 
solve nonlinear least-squares problems. Finally, 
we present and discuss numerical results in 
Section~\ref{sec:results}.

\section{Approach}
\label{sec:approach}

\subsection{Large kinetic networks}

A major topic in systems biology is the study of the dynamical evolution
of bio-chemical mechanisms within a well-defined, biology-related
context. The bio-chemical mechanisms in such a compound under consideration
are typically given as a, possibly huge, set of chemical reactions
between numerous species forming a large kinetic network.
Assuming the general principle of mass action kinetics, this large
network transforms readily to a system of $n$ ordinary differential
equations (ODEs) leading to an autonomous initial value problem (IVP) 
\begin{equation}
  y' = f(y\,;\, p), \qquad y(t_{0}) = y_{0}, \qquad p \in \mathbb{R}^q 
\label{eq:IVP}
\end{equation}
where the rate of change in the species vector, $y' \in \mathbb{R}^n$, 
is described by the term on the right-hand side, $f(y;p)$, depending 
on both the species, $y \in \mathbb{R}^n$, and the parameter vector, 
$p \in \mathbb{R}^{q}$.  The initial condition vector, $y_{0}$, has 
the same dimension as the species vector $y$.  In BioPARKIN, the ODE 
systems are solved numerically with LIMEX, a linearly implicit Euler method 
with extrapolation that is especially suited for stiff differential 
equations \citep{limex,limex2,Ehr99}.  LIMEX is a numerical integrator with 
adaptive stepsize control that allows for a computation of the solution $y$ 
at arbitrary time points with prescribed accuracy by using an appropriate 
interpolation scheme.  This is often not possible with other ODE solvers.  
LIMEX can be applied to differential-algebraic equations as well, which 
allows for the processing of algebraic constraints in BioPARKIN.

It is assumed that some discrete experimental data (in form of species 
concentrations versus time),
\begin{equation}
    (\tau_{1},z_{1}),\,\ldots,\,(\tau_{M},z_{M}),
\end{equation}
are available.  Note that frequently only a certain amount of the $n$
species concentrations are measurable observables, if at all. The task
at hand now reduces to quantify the $q$ unknown components of the
parameter vector, $p$, by comparison between computed model values 
and measured data. 

A complete data set, of course, must include prescribed statistical
tolerances, $\delta z_{j} \, (j = 1, \ldots, M)$, for each measurement
as well. The mathematically correct handling of these will be described
in Section~\ref{sub:Parameter-identification}.

\paragraph{Breakpoint handling.}
A sudden event (maybe from outside the biological system) is handled
by introducing a breakpoint, $t_{b} > t_{0},$ and subsequently, splitting
the ODE system into a $y^{-}$-part for $t_{0} < t \leq t_{b}$, and
a $y^{+}$-part for $t_{b} < t$, 
\begin{eqnarray}
  (y^{-})' & = & f(y^{-}; \, p),\qquad y^{-}(t_{0}) = y_{0} \\
  (y^{+})' & = & f(y^{+}; \, p),\qquad y^{+}(t_{b}) = g(y^{-}(t_{b}) \, ; \, p)
\end{eqnarray}
where $g:\mathbb{R}^{n} \times \mathbb{R}^{q} \longrightarrow \mathbb{R}^{n}$
is a mapping of the initial conditions, possibly dependent on the
parameter vector, $p$. Note that, in BioPARKIN, breakpoints have to 
be defined beforehand and hence, they must be independent of the time 
course of $y$.  This approach of splitting the ODE system with respect 
to time particularly applies in case of multiple experiments.

In SBML such breakpoints are defined via ``events'' with trigger expressions 
in the form
\begin{center} $\mathtt{ eq(time,t_b) }.$ \end{center}
Many other present simulation tools cannot handle this kind of event because 
the numerical integrator simply does not stop at time $t_b$.

\paragraph{Multiple experiments.}
The design of experiments almost always includes different conditions
such that the effects of these different conditions on the system
under investigation can be observed and studied. In the simplest case,
calibration measurements might be necessary, for example, or data related 
to different initial conditions, $y_{0,1}, y_{0,2}, \ldots,y_{0,\nu}, \ldots$,
are given. Numerically, these situations can be handled by the concatenation
of several IVPs,
\begin{equation}
  y_{\nu}' = f_{\nu}(y_{\nu} \, ; \, p), \qquad 
  y_{\nu}(t_{0,\nu}) = y_{0,\nu}, \quad 
  \nu = 1,2,\ldots,
\end{equation}
very similar to the management of breakpoints/events.
If required, the solution $y_{\nu}$ corresponding to the (virtual)
initial timepoint, $t_{0,\nu}$, can readily be shifted to the (original)
initial time, $t_{0}$, for comparison or plotting purposes.

\subsection{Parameter identification}
\label{sub:Parameter-identification}

Following the fundamental idea of Gauss, parameter identification
is, as implemented in BioPARKIN, equivalent to solving the \emph{weighted}
least-squares problem,
\begin{equation}
 \frac{1}{M} \sum_{j=1}^{M} \| D_{j}^{-1} \big(y(\tau_{j}\,;\, p) - z_{j}\big) \|_{2}^{2} = \min, 
 \label{eq:Fmin}
\end{equation}
with diagonal weighting $(n,n)$-matrices,
\begin{equation}
 D_{j} := \mathrm{diag} \big( (\delta z_{j})_{1},\,\ldots,\,(\delta z_{j})_{n} \big),
 \qquad j = 1, \ldots, M.
\end{equation}
Note that, if not all components of a datum, $z_{j} \in \mathbb{R}^{n}$,
are available at a specific measurement time point, $\tau_{j}$,
then the missing data in the least-squares formulation is simply replaced
by the computable model value, therefore effectively neglecting the
corresponding contribution in the sum of Equation~(\ref{eq:Fmin}). 
The corresponding entry in $D_{j}$ is then set to one. 

If, on the other hand, a component of given error tolerance, $\delta z_{j}$,
or even the whole vector, is put to zero, this contribution to the
sum in Equation~(\ref{eq:Fmin}) is also taken out, and considered as a 
(nonlinear) equality constraint to the least-squares formulation instead.

In the (hopefully rare) case of missing error tolerances in the data file, 
the measurement tolerances are simply set to 
\begin{equation}
  \big( \delta z_{j} \big)_{\ell} = \max \big\{ \big| \big(z_{j}\big)_{\ell} \big|, \,
                                                \mathrm{thres}(y_{\ell}) \big\},\quad 
  \ell = 1, \ldots, n,
\end{equation}
with some user specified threshold mapping, $\mathrm{thres}(y_{\ell})$. 
If this threshold value is not defined, it is set to zero.

The least-squares problem (\ref{eq:Fmin}) may be written 
even more compactly as
\begin{equation}
  \|F(p)\|_{2}^{2} \equiv F(p)^{^{T}}F(p) = \min,
\end{equation}
where $F \, : \, \mathbb{R}^{q} \rightarrow \mathbb{R^{L}}$ 
is a nonlinear mapping and structured as a stacked vector of 
length $L = nM$,
\begin{equation}
  F(p) = \left[ \begin{array}{c}
                  D_{1}^{-1} \big( y(\tau_{1} \, ; \, p) - z_{1} \big) \\
                  \vdots \\
                  D_{M}^{-1} \big( y(\tau_{M}\, ; \, p) - z_{M} \big)
                \end{array}\right].
\end{equation}
If \emph{not} \emph{all} components of a measurement, $z_{j}$, 
are given, the number $L$ will accordingly be smaller, $L < nM$.

\subsection{Parameter constraints}

In order to enforce constraints such as positivity or upper and lower
bounds on the unknown parameters to be determined in the model, a (differentiable)
transformation, $\varphi:\mathbb{R}^{q}\longrightarrow\mathbb{R}^{q}$,
can be introduced resulting in a different parametrisation, $u$, of
the model ODE system,
\begin{equation}
 p = \varphi(u),\quad y' = f(y\,;\,\varphi(u)) = \tilde{f}(y\,;\, u)
\end{equation}
A global positivity constraint on the parameter vector, $p$, can be 
achieved, for example, by the (component-wise) exponential transformation
\begin{equation}
  p_{i} = \exp(u_{i}),\qquad i=1,\ldots,q
\end{equation}
To impose an upper and a lower bound, $A$ and $B$, respectively,
a sinusoidal transformation 
\begin{equation}
  p_{i} = A + \frac{B-A}{2}\,\left(1 + \sin u_{i}\right),\qquad i=1,\ldots,q
\end{equation}
can be used. For a single bound, $C$, as last example in this section,
a root square transformation
\begin{equation}
  p_{i} = C\pm\left(1 - \sqrt{1 + u_{i}^{2}}\right),\qquad i=1,\ldots,q
\end{equation}
(with the upper sign for an upper bound and the lower sign for a lower
bound) is possible. 

The last two transformation formulas are especially eligible since, at least 
for small perturbations $\mathrm{d}p_{i} \approx \varphi' \, \mathrm{d}u_{i}$,
the differentials are bounded and, most importantly, are essentially independent 
of the new parametrisation, $u$.

Note that the application of any transformation of the parameters
obviously changes the sensitivities of the parameters to the dynamical
evolution of the ODE system. Therefore, it is strongly recommended 
that parameter constraints should only be applied in order to prevent 
the parameter vector components, $p_{i}$, from taking on physically 
meaningless values. The better choice in this case would be to change 
the model equations since model and data seem to be incompatible.

\subsection{Parameter scaling}

In general, a scaling-invariant algorithm, i.e.~an algorithm that is 
invariant under the choice of units in a given problem, is (almost) 
indispensable to guarantee any reliable results.  Therefore, the following 
scaling strategy within the course of the Gauss-Newton iteration has been 
implemented:  in each iteration step $k$, an internal weighting vector, 
$pw\in\mathbb{R}^{q}$, is used to define local scaling matrices, 
$W_{k}$ , by
\begin{equation}
  W_{k} = \mathrm{diag}(pw_{1}, \ldots, pw_{q})
\end{equation}
with locally given
\begin{equation}
  pw_{i} := \max \left\{ |p_{i}^{k}|,\:\mathrm{thresh}(p_{i}) \right\}, \qquad 
  i = 1, \ldots, q
\end{equation}
where $p_i^k$ are the current parameter iterates, and 
${\mathrm{thresh}(p_{i})>0}$ are suitable threshold values 
for scaling chosen by the user. Consequently, any relative 
precision of parameter values below these prescribed threshold 
values will be meaningless.



\section{Methods}
\label{sec:methods}

\subsection{Affine covariant Gauss-Newton algorithm}
\label{sub:Gauss-Newton-Method}

Starting with an initial guess, $p^{0} \in \mathbb{R}^{q}$, 
the (damped) Gauss-Newton method is given as
\begin{equation}
  p^{k+1} = p^{k} + \lambda_{k}\Delta p^{k},\qquad k = 0, 1, \ldots
\end{equation}
Here, the step-length, $0< \lambda_{k} \le 1$, is recomputed successively
in each iteration (see below). The update, $\Delta p^{k}$, is the
minimum norm solution to the \emph{linear} least-squares problem,
\begin{equation}
  \| F'(p^{k})\,\Delta p^{k} + F(p^{k}) \| \overset{!}{=} \min.
\end{equation}
The $(L\times q)$-Jacobian matrix, $F'(\cdot)$, can be approximated
by stacking the rows of the sensitivity matrices, $S(\tau_{j})$, corresponding
to the measurement points $(\tau_{j},\, z_{j})$,
\begin{equation}
  J = \left[\begin{array}{c}
              D_{1}^{-1} S(\tau_{1}) \\
              \vdots\\
              D_{M}^{-1} S(\tau_{M})
            \end{array} \right].
\end{equation}
Herein the sensitivity matrices, $S(\tau_j)$, are samples of the solution
trajectories of the inhomogeneous \emph{variational equation}
\begin{equation}
       S' = f_{y}\Big(y(t\,;\,p^{k})\,;\,p^{k}\Big) S + 
            f_{p}\Big(y(t\,;\,p^{k})\,;\,p^{k}\Big), \quad 
   S(t_0) = 0
\end{equation} 
taken at the measurement time points, $\tau_{j}$. The terms $f_y$ and $f_p$ 
on the right hand side are computed analytically by symbolic differentiation.
The variational equation is solved simultaneously with the IVP (\ref{eq:IVP}), 
representing an ODE system of $((n+1)\times q)$ equations in total. To avoid 
expensive factorisations of the iteration matrix within LIMEX, it is replaced 
by its block-diagonal part, as proposed in \citep{Schlegel04}. The linearly-implicit 
extrapolation algorithm allows such an approximation, as long as the characteristics 
of the dynamic system are preserved, which is satisfied here. By using this sparsing, 
the effort required for sensitivity evaluation does not grow quadratically with the 
number of parameters, $q$, but only linearly. Hence, reasonable computing times are 
achieved (compare also Table~\ref{Tab:01}).

For reasons of comparison with other software tools, the Jacobian matrix 
can alternatively be approximated by computing the difference quotient, for 
$\ell = 1, \ldots, L$ and $i = 1, \ldots, q$,
\begin{equation}
  J_{\ell,i} = \frac{1}{h} \Big(F_{\ell}(p+e_{i}h) - F_{\ell}(p)\Big), \qquad 
  h = \mathcal{O}\Big( \left|p_{i}\right| \cdot \sqrt{\mathrm{eps}} \Big),
\end{equation}   
whereby $\mathrm{eps}$ it the relative machine precision. In BioPARKIN, the user can 
optionally invoke a feedback strategy in which the finite difference disturbance is 
additionally adapted to the current values of $F_{\ell}$. 

All approaches to compute the Jacobian matrix make sure that, at each current parameter 
estimation, $p^{k}$, the approximation $J\approx F'(p^{k})$ is valid.  Note, however, 
that the Jacobian computed by numerical differentiation is generally less accurate than 
the Jacobian obtained via the variational equation.

In passing, the notation of the so-called \emph{simplified} \emph{Gauss-Newton}
\emph{correction}, $\overline{\Delta p}^{k+1}$, as the minimum norm
solution to
\begin{equation}
  \|J(p^{k})\,\overline{\Delta p}^{k+1} + F(p^{k+1})\| \overset{!}{=} \min,
\end{equation}
may also be introduced for later use.

\subsection{Threshold-related scaling}

Often, model species and model parameters cover a broad range of 
physical units and their values can vary over orders of magnitude. 
To achieve comparability, the sensitivity values have to be 
normalised by the absolute values of species and parameters 
to obtain scaled sensitivity matrices,
\begin{equation}
  S_{ij}(t) = \left( \frac{\partial y_{i}}{\partial p_{j}}\right)(t) \cdot 
              \frac{ \max \{ |p_{j}| \, , \, \mathrm{thres}(p_{j}) \} }
                   { \max \{ \max\limits_{t\in I} |y_{i}(t)| \, , \, \mathrm{thres}(y_{i}) \} }
\end{equation}
where $\mathrm{thres}(\cdot)$ are user-specified threshold values
for parameters and species, respectively, and the integration time
interval of the ODE system, $I$, is used. In BioPARKIN, the absolute 
values of these scaled sensitivity values are displayed 
(see Figure~\ref{fig:02} as an example).

\subsection{Subcondition monitor}
\label{sub:Subcondition-Monitor}

For the solution of the linear least-squares problem in each iteration
step, a QR-decomposition of the associated Jacobian $(L,q)$-matrix,
$J = F'(p)$, 
\begin{equation}
  Q\, J\,\Pi = \left(
                    \begin{array}{c}
                          R\\
                          0
                    \end{array}
                \right)
\end{equation}
by applying Householder reflections with additional column pivoting
is realised in BioPARKIN. Here, for simplicity, the full rank case
is assumed where $q \leq L$ and $R$ is an upper triangular $(q,q)$-matrix,
$R = (r_{ij})$. The permutation, $\Pi$, is determined such that
\begin{equation}
  |r_{11}| \geq |r_{22}| \geq \ldots \geq |r_{qq}|.
\end{equation}
For some required accuracy, $\delta > 0$, given by the user, the 
\emph{numerical rank}, $\ell := \mathrm{rnk}(J)$, indispensable to 
the successful solution of ill-posed problems, is then defined by 
the inequality
\begin{equation}
  |r_{\ell+1,\ell+1}| < \delta \, |r_{11}| \leq |r_{\ell\ell}|.
\end{equation}
In general, the maximum of all given measurement tolerances, $\delta z_j$, 
is a suitable choice for this accuracy, 
$\delta := \max\limits_{i,j} \{ (\delta z_{j})_{i} \}$. 
In BioPARKIN, however, this choice is left to the user, who has 
to specify a tolerance XTOL. This tolerance is assigned to $\delta$.

Note that this definition of the numerical rank is highly biased by both 
row and column scaling of the Jacobian. Introducing, nevertheless, the 
so-called subcondition number, for $\ell = q$, by 
\begin{equation}
  \mathrm{sc}(J) := \frac{|r_{11}|}{|r_{qq}|} \leq \mathrm{cond}_{2}(J),
\end{equation}
it follows that, if $\delta \cdot \mathrm{sc}(J) \geq 1$, 
the Jacobian will certainly be rank-deficient. In this case, 
a rank-deficient pseudo-inverse is realised in BioPARKIN, 
either by a QR-Cholesky variant or by a QR-Moore-Penrose variant \citep{DdSa:80}. 
Both cases of pseudo-inverses of the Jacobian, $J$, will be denoted 
by $\left( J^{\ell} \right)^{+}$.

\subsection{Steplength strategy}

In order to determine an optimal damping parameter, $0 < \lambda_{k} \leq 1$, 
in each Gauss-Newton step, a first estimate $\lambda_{k}^{(0)}$ is calculated 
in BioPARKIN from a theoretical prediction on the basis of the former iterate step,
\begin{equation}
   \begin{array}{rcl}
     \lambda_{k}^{(0)} &  = & \min \{ 1, \mu_{k} \} \\[2ex]
               \mu_{k} & := & \left[ \|\Delta p^{k-1}\| \, \|\overline{\Delta p}^{k}\| \, / \,
                                   ( \, \rho_{k} \, \|\Delta p^{k}\| \, )
                            \right] \cdot \lambda_{k-1} \\[2ex]
              \rho_{k} & := & \left\| \left[ I_q - J(p^{k})^{+} J(p^{k-1}) \right] \overline{\Delta p}^{k} \right\| \, .
   \end{array}
\end{equation}

If this first \emph{a priori} estimate, $\lambda_{k}^{(0)}$, 
fails in the \emph{natural monotonicity test},
\begin{equation}
  \left\| \overline{\Delta p}^{k+1} \right\| < \left\| \Delta p^{k} \right\| \, ,
\end{equation}
then an additional \emph{correction strategy} is invoked to compute the
\emph{a posteriori} estimates,
\begin{equation}
    \lambda_{k}^{(\nu)} = \min \left\{ 1 \, , \,
                                       \frac{1}{2}\lambda_{k}^{(\nu-1)} , \ ,
                                       \frac{1}{2}\mu_{k}^{(\nu-1)}
                               \right\} ,
    \qquad \nu = 1, 2, \ldots
\end{equation}
where
\begin{equation}
   \mu_{k}^{(\nu-1)} := 
     \frac{\|\Delta p^{k}\|}{\|\overline{\Delta p}^{k+1,\nu-1} - (1-\lambda_{k}^{(\nu-1)})\,\Delta p^{k}\|}
     \cdot (\lambda_{k}^{(\nu-1)})^{2} \, .
\end{equation}
For details see \citep{Deu04} and \citep{Die11}. 

As experience shows, the \emph{a posteriori} loop is rarely activated. 
To avoid an infinite loop, however, it is ensured that both estimates, 
$\lambda_{k}^{(0)}$ and $\lambda_{k}^{(\nu)}$, $\nu=1,2,\ldots$,
always satisfy the condition
\begin{equation}
\lambda_{k}^{(\nu)}\geq\lambda_{\mathrm{min}},\qquad\nu=0,1,2,\ldots
\end{equation}
with a minimal permitted damping factor, $\lambda_{\mathrm{min}}$,
provided by the user. In case $\lambda_{k}^{(\nu)}<\lambda_{\mathrm{min}}$
deliberate rank reduction is invoked, which usually leads to larger
damping factors. Otherwise, the Gauss-Newton iteration will be stopped.

\subsection{Deliberate rank reduction}

A deliberate rank reduction may additionally help to avoid an iteration
towards an attractive point, $\hat{p}$, where the associated Jacobian
matrix, $J(\hat{p})$, becomes singular. The general idea of this
device is to reduce the maximum permitted rank in the $QR$ decomposition
until the natural monotonicity will be fulfilled again or, of course,
no further rank reduction is possible. The subroutine as implemented 
in BioPARKIN is as follows.

To start with, let $q$ denote the current rank. The ordinary Newton
correction, $\Delta p^{k}$, is then recomputed with a prescribed
maximum allowed rank, $\ell = q - 1$. With the new (trial) correction,
$\Delta p^{k,\ell}$, a new \emph{a priori} damping factor, a new
trail iterate, and a new simplified correction,
\begin{eqnarray}
           \lambda_{k}^{(0,\ell)} & = & \min\left\{ 1,\,\mu_{k}^{(0,\ell)}\right\} ,\\
                     p^{(0,\ell)} & = & p^{k} + \lambda_{k}^{(0,\ell)}\Delta p^{k,\ell},\\
                \Delta p^{k,\ell} & = & -J^{\ell}(p^{k})^{+} F(p^{k}),\\
 \overline{\Delta p}{}^{(0,\ell)} & = & -J^{\ell}(p^{k})^{+} F(p^{(0,\ell)}),
\end{eqnarray}
are computed, respectively. 

If now the monotonicity test is successfully passed, the 
Gauss-Newton iteration proceeds as usual. Otherwise, the damping factors, 
$\lambda_{k}^{(\nu,\ell)}(\nu=1,2,\ldots)$, are calculated using the 
\emph{a posteriori} estimates as given above.  If in the \emph{a posteriori} 
loop, in turn, $\lambda_{k}^{(\nu,\ell)}<\lambda_{\mathrm{min}}$ occurs, 
the maximum allowed rank is further lowered by one and, again, the repetition 
of the rank reduction step starts once more. 

This rank reduction procedure is carried out until natural monotonicity,
$\|\overline{\Delta p}^{(\nu,\ell)}\| \leq \|\Delta p^{k,\ell}\|$,
holds true or, alternatively, a final termination criterion, 
$\ell < \ell_{\mathrm{min}} \, (0 < \ell_{\mathrm{min}} < q)$, is reached. 

Note that an emergency rank reduction can occur in a step where the
rank of the Jacobian, $J(p^{k})$, has already been reduced because
of the subcondition criterion.

\subsection{Convergence}

\begin{table}[!t]
\caption{Typical protocol of parameter identification run with full data, 
         here for the model EpoRcptr (cf.~Section~\ref{sub:EpoRcptr}).\label{Tab:02}}
\begin{indented}
\item[]
\begin{tabular}{@{}crcrrr}\br
             G-N It. &           Normf  &     &       Normx  &  Damp.~Fctr. &  Rank  \\ \mr
               0     &   4.1941414e+01  &     &   2.115e-02  &              &     6  \\
               1     &   4.1936708e+01  &  *  &   2.094e-02  &     0.01000  &        \\
               1     &   4.1936708e+01  &     &   2.469e-02  &              &     6  \\
               2     &   4.1751843e+01  &  *  &   1.669e-02  &     0.41932  &        \\
               2     &   4.1751843e+01  &     &   3.373e-02  &              &     6  \\
               3     &   4.1655239e+01  &  *  &   2.266e-02  &     0.42693  &        \\
               3     &   4.1655239e+01  &     &   1.024e-01  &              &     6  \\
               4     &   4.1639220e+01  &  *  &   7.410e-02  &     0.19117  &        \\
               4     &   4.1639220e+01  &     &   1.076e-01  &              &     6  \\
               5     &   4.1631470e+01  &  *  &   4.854e-02  &     0.37178  &        \\
               5     &   4.1631470e+01  &     &   1.538e-02  &              &     6  \\
               6     &   4.1547355e+01  &  *  &   1.816e-03  &     1.00000  &        \\
               6     &  \multicolumn{4}{r}{incompatibility factor: 0.14248} &        \\
               6     &   4.1547355e+01  &     &   6.366e-03  &              &     6  \\
               7     &   4.1542667e+01  &  *  &   2.140e-04  &     1.00000  &        \\
               7     &  \multicolumn{4}{r}{incompatibility factor: 0.42707} &        \\
               7     &   4.1542667e+01  &     &   3.339e-05  &              &     6  \\
               8     &   4.1542118e+01  &  .  &   1.783e-08  &     1.00000  &        \\
               8     &  \multicolumn{4}{r}{incompatibility factor: 0.00526} &        \\ \br
\end{tabular}
\item[] Requested identification accuracy has been $\mathrm{xtol} = 10^{-4}$. 
        A star in the third column indicates values corresponding to 
        simplified Gauss-Newton corrections.
\end{indented}
\end{table}

%
%

As the solution $p^{\ast}$ is approached, the Gauss-Newton method converges 
linearly with an asymptotic convergence factor $\kappa(p^{\ast})$.  This 
quantity $\kappa$, called {\em incompatibility factor}, is monitored by NLSCON 
and must be smaller than 1 to obtain convergence.  Problems that satisfy this condition 
are called {\em adequate} problems.  If model and measurement values match exactly, 
i.e. $F(p^{\ast}) = 0$, then $\kappa(p^{\ast}) = 0$ and the method converges 
quadratically just as Newton's method.  This so-called compatible case, however, 
does not occur in practice since experimental measurements are never exact.  
For inadequate nonlinear least-squares problems, the adaptive damping strategy 
will typically yield values $\lambda_k \approx 1/\kappa < 1$, and too small 
damping factors result in fail of convergence.  Vice versa, this effect can be 
conveniently taken as indication of the inadequacy of the inverse problem 
under consideration \citep{Deu04}.  In this case, model equations or the 
initial parameter guess $p^0$ should be changed.  A typical NLSCON output 
protocol in case of successful convergence is shown in Table~\ref{Tab:02}. 
In the convergent phase, the damping factors approach 1 and finally $\kappa<1$.



\section{Results of numerical experiments}
\label{sec:results}

\begin{table}[!t]
\caption{Comparison of computing times w.r.t.~different models.\label{Tab:01}}
\begin{indented}
\item[]
\begin{tabular}{@{}rrrrr}\br
                                 &  {GynCycle}  &  {BovCycle}  &  {BIOMD008} & {EpoRcptr} \\ \mr
 \multicolumn{1}{l}{Model Characteristics}    &            &          &        \\
                       \#Species &       33   &       15   &      5   &    7   \\
                    \#Parameters &      114   &       60   &     21   &    9   \\
                     \#Reactions &       54   &       28   &     13   &    9   \\
                                 &            &            &          &        \\
  \multicolumn{1}{l}{Simulation} &            &            &          &        \\
  \ \ \ \ \ BioPARKIN${}_{0}$ (adpt. $h$) &      3.2s  &      0.8s  &    0.1s  &  0.1s  \\
  COPASI${}_{1}$ ($h = 10^{-2}$) &      1.4s  &      0.6s  &    0.2s  &  0.2s  \\
  COPASI${}_{2}$ ($h = 10^{-3}$) &      7.2s  &      4.0s  &    1.4s  &  1.1s  \\
                                 &            &            &          &        \\
 \multicolumn{1}{l}{Sensitivity} &            &            &          &        \\
            BioPARKIN            &            &            &          &        \\  
     $(*)$\ (var.~eq., overview) &       49s  &     12.9s  &    0.7s  &  1.7s  \\
            (var.~eq., overview) &      117s  &     29.2s  &    0.9s  &  2.0s  \\
          (num.~diff., overview) &      309s  &     35.4s  &    0.8s  &  0.2s  \\
                                 &            &            &          &        \\
    COPASI${}_{1}$ (grand total) &       94s  &     18.1s  &    1.0s  &  0.3s  \\
    COPASI${}_{2}$ (grand total) &      328s  &    115.6s  &    8.3s  &  1.5s  \\ \br
\end{tabular}
\item[] Benchmark times are rounded to one decimal. Integration was done in [0,100] with 
        time units [s] or [d], accordingly.  For comparison reasons,
        $\mathtt{rtol} = 10^{-6}$ and $\mathtt{atol} = 10^{-12}$ have been used
        in all rows (except $(*)$) as accuracy for the ODE solvers. 
        COPASI run times have been measured by batch processing, {\em ex\-cluding} the time 
        spent for file I/O.
\item[] In COPASI, sensitivities were computed by numerical differentiation.
        In BioPARKIN, sensitivities were computed by either solving the variational equation 
        (var.~eq.) or by numerical differentiation (num.~diff.). In a sensitivity {\em overview}, 
        sensitivities are plotted over the complete time interval (for an example, see 
        Figure~\ref{fig:02}).
\item[] $(*)$ Var.~Eq.~computing times: values have been achieve with slightly lower but 
        still more than sufficient accuracy ($\mathtt{rtol}=10^{-5}$, $\mathtt{atol}=10^{-7}$). 
\end{indented}
\end{table}

\begin{figure}[!tpb]
\centerline{\includegraphics[scale=0.42]{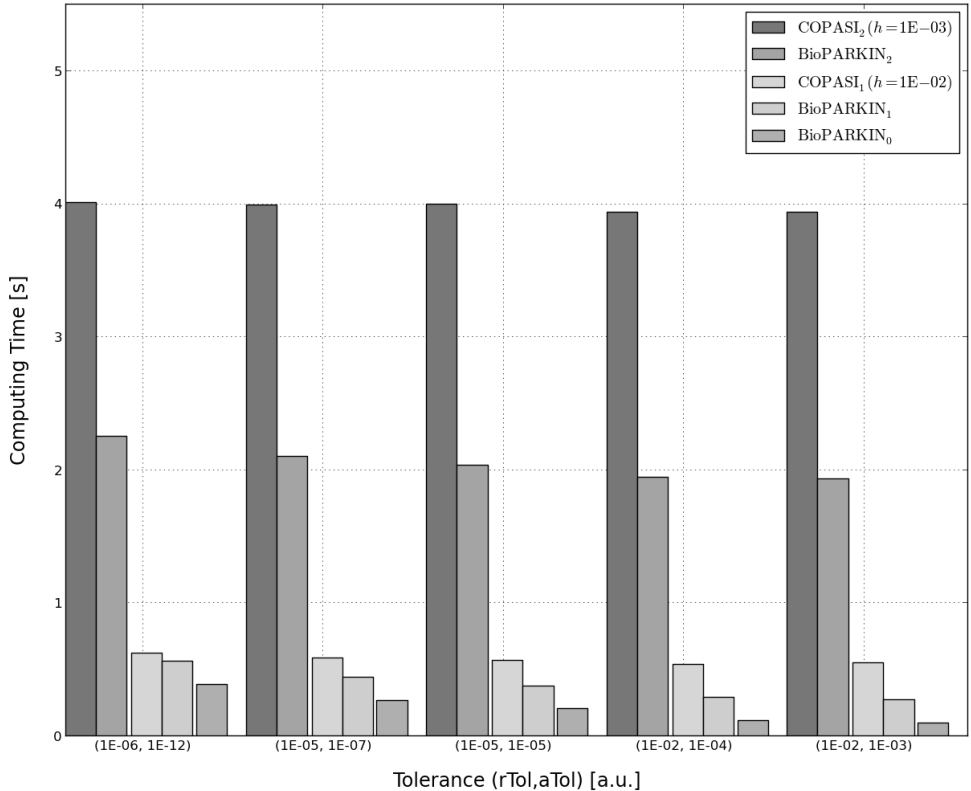}}
\caption{BovCycle: Computing times for the BovCycle model w.r.t.~different integrator tolerances.
         The cases BioPARKIN${}_{1}$ and BioPARKIN${}_{2}$ are interpolating at exactly as many 
         sample points as requested for the COPASI${}_{j}$ $(j=1,2)$ cases, respectively, in 
         addition to the adaptive time points.  Note that these artificially high numbers of 
         sample points are unusual and absolutely unnecessary for trajectory computations with 
         BioPARKIN and that, for comparision reasons only, these numbers have been applied here.  
         Additionaly, BioPARKIN${}_{0}$ denotes the timings in case of no interpolation at all.}\label{fig:_1}
\end{figure}

\begin{figure}[!tpb]
\centerline{\includegraphics[scale=0.42]{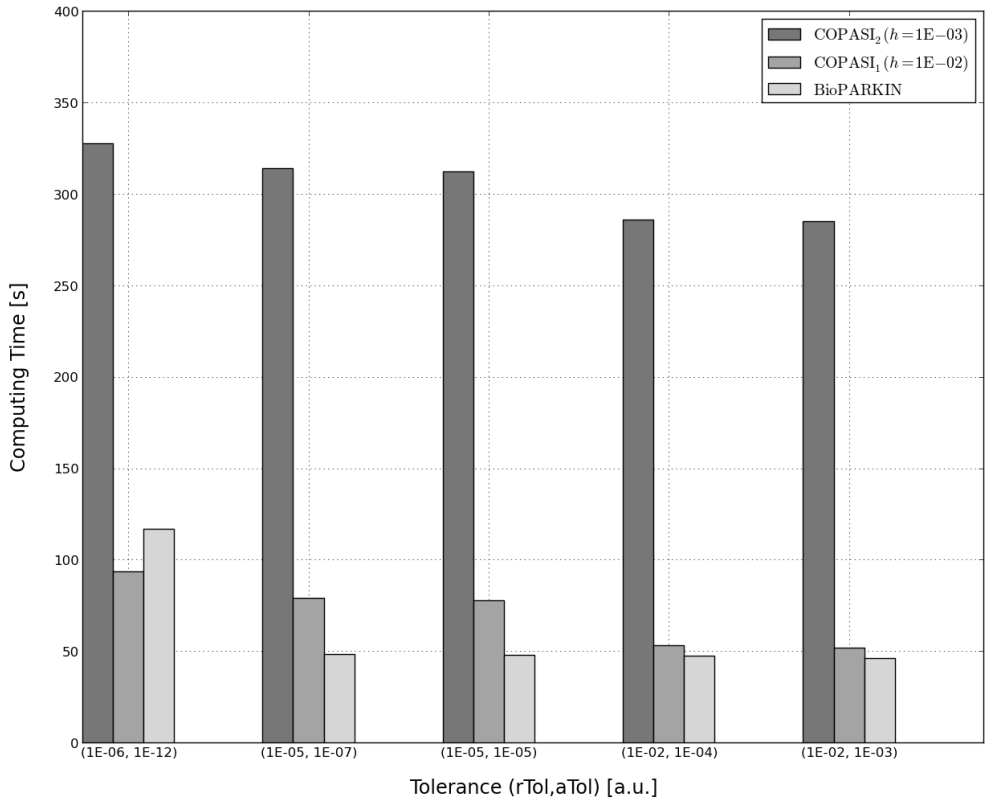}}
\caption{GynCycle: Computing times for the variational equation w.r.t.~different integrator tolerances. 
         Note that BioPARKIN integrates the variational equation system while COPASI takes 
         finite differences for the computation.}\label{fig:00}
\end{figure}

\begin{figure}[!tpb]
\centerline{\includegraphics[scale=0.6]{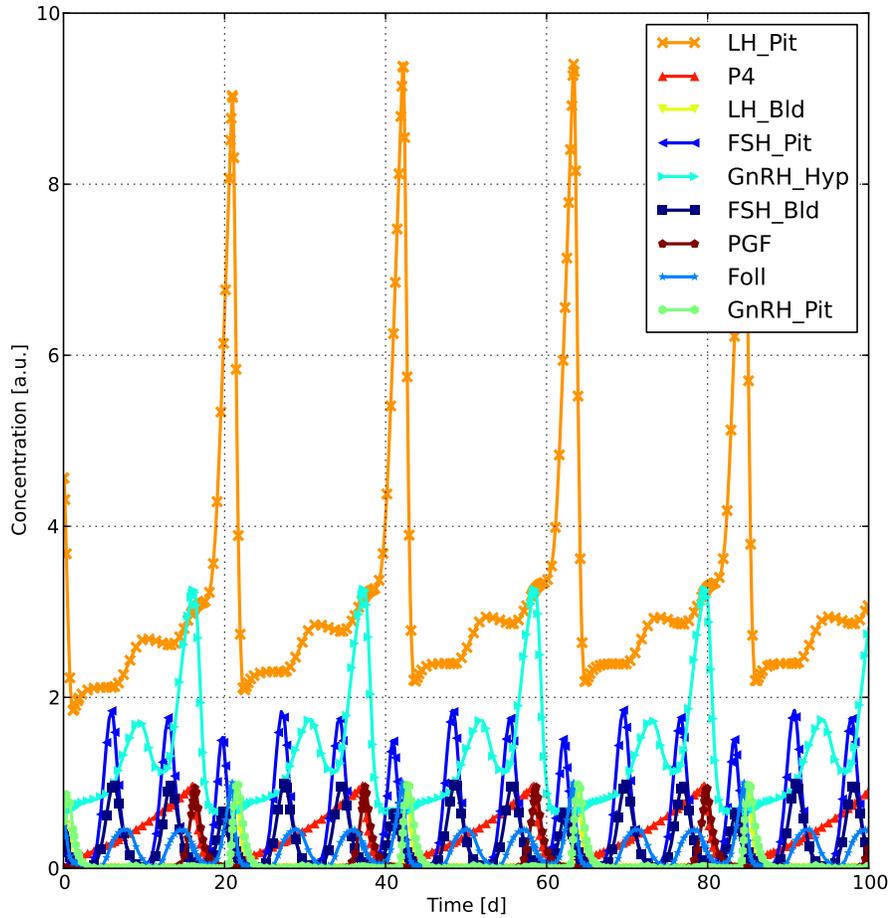}}
\caption{BovCycle: Trajectories of model simulation of selected species.}\label{fig:01}
\end{figure}

\begin{figure}[!tpb]
\centerline{\includegraphics[scale=0.6]{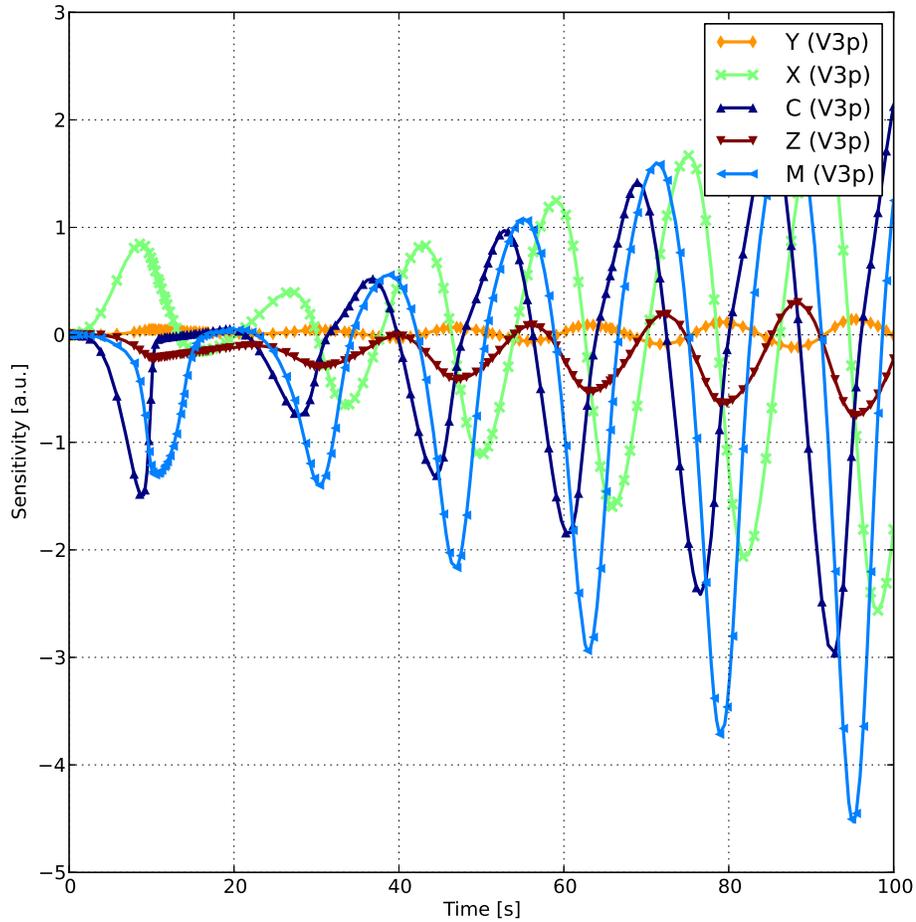}}
\caption{BIOMD008: Sensitivity trajectories of the variational equation w.r.t.~parameter V3p.}\label{fig:02}
\end{figure}


\begin{figure}[!tpb]
\centerline{\includegraphics[scale=0.6]{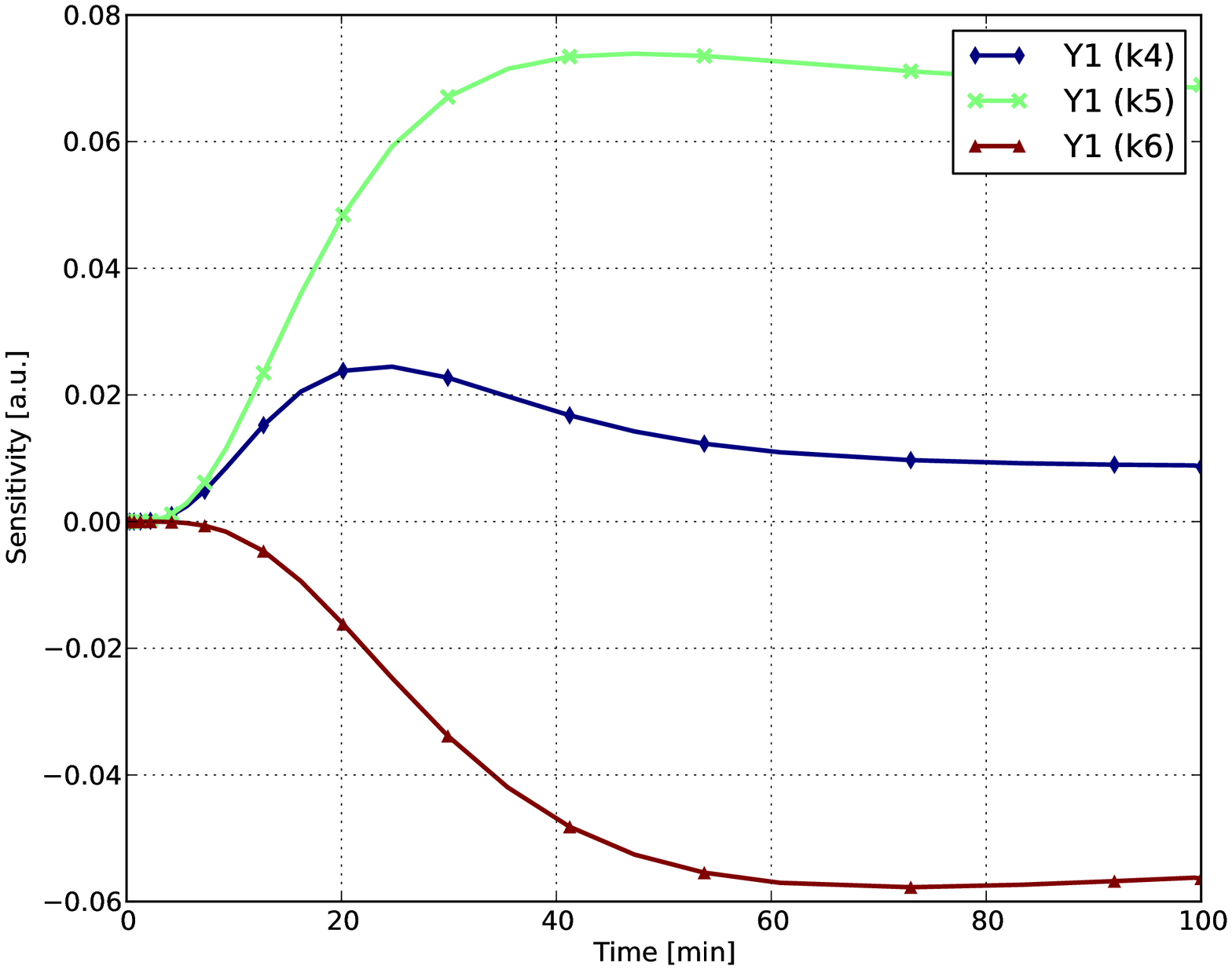}}
\caption{EpoRcptr: Sensitivity trajectories of measurement variable $Y_1$.}\label{fig:04}
\end{figure}

\begin{figure}[!tpb]
\centerline{\includegraphics[scale=0.6]{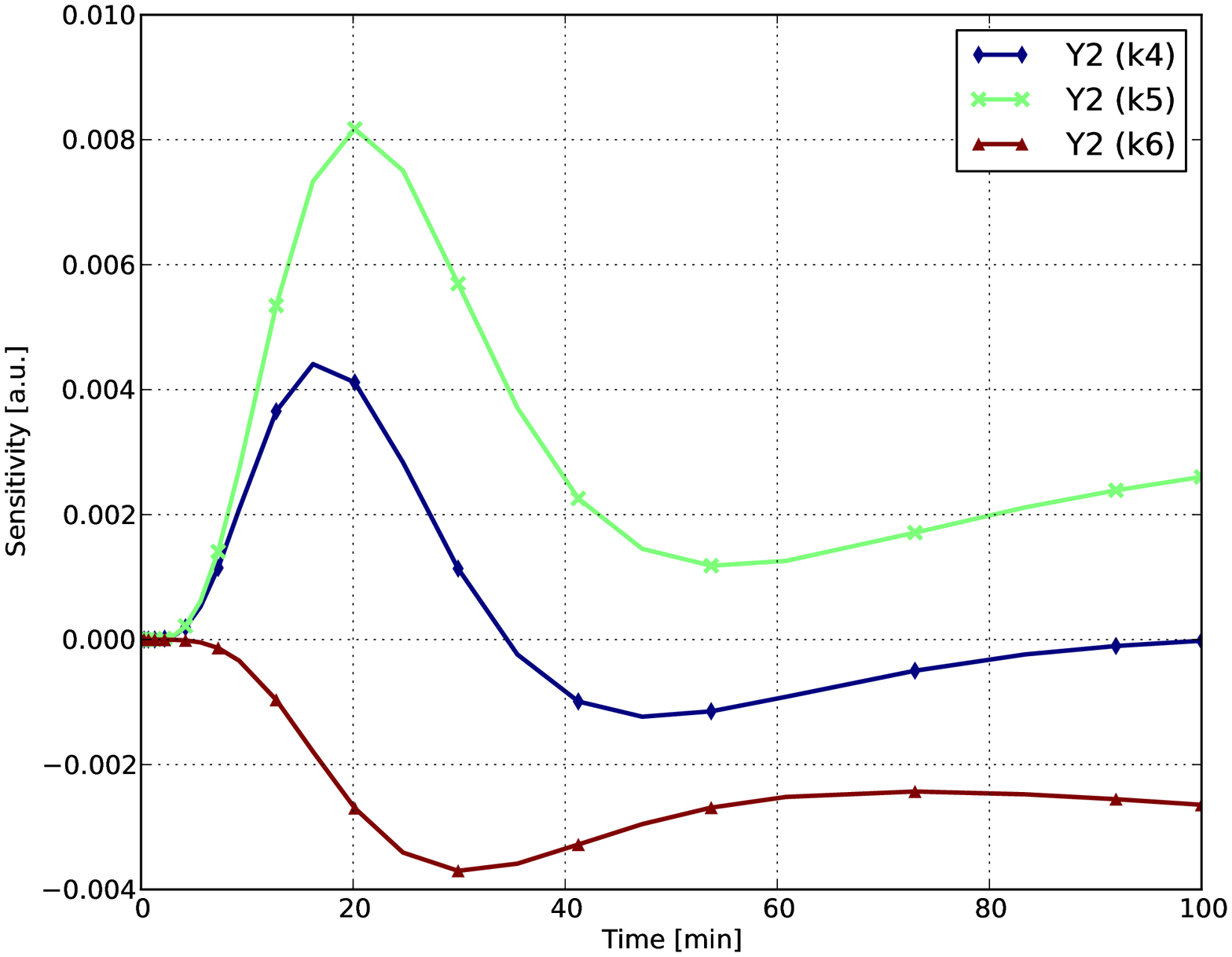}}
\caption{EpoRcptr: Sensitivity trajectories of measurement variable $Y_2$.}\label{fig:05}
\end{figure}

\begin{figure}[!tpb]
\centerline{\includegraphics[scale=0.6]{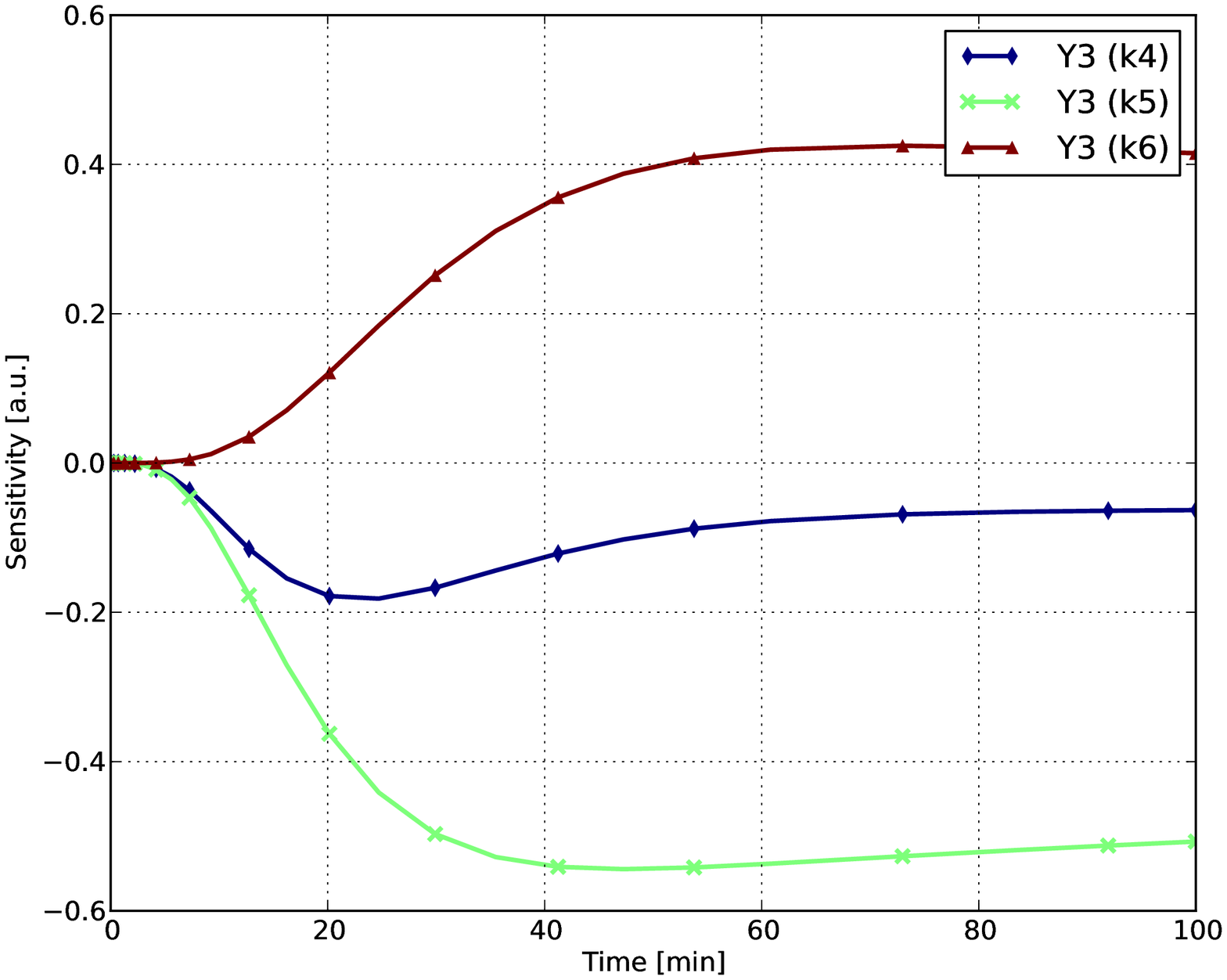}}
\caption{EpoRcptr: Sensitivity trajectories of measurement variable $Y_3$.}\label{fig:06}
\end{figure}

This section illustrates the use of BioPARKIN and PARKINcpp with actual models. First, 
two models developed by the Computational Systems Biology group at Zuse Institute Berlin 
are presented.  Next, a third model was obtained from the BioModels database, a website 
with curated SBML models \citep{le2006biomodels}. And last but not least, a variant of 
a EPO receptor model from the same database is taken, as it was already published in \cite{Hengl07}.  
All subsequent computations have been performed on an Intel Core 2 Dual CPU (T7200 @ 2.0GHz).
In addition, for comparison reason, all forward simulations have been repeated by using 
COPASI \citep{hoops2006copasi}.  Note that the stiff ODE solver LSODAR \citep{lsoda0,lsoda1} 
is used in COPASI in contrast to LIMEX.  In fact, it seems that, for the computation of 
any model trajectories, the researcher is forced to supply an equidistant time grid in COPASI.  
Thereby, the accuracy of the ODE solution, as set by the user in the values $\mathtt{atol}$ 
and $\mathtt{rtol}$, can easily be foiled in the sense that essential details of model
trajectories are simply neglegted in COPASI if the chosen equidistant time grid happens 
to be too coarse.  Note that this is surely not contradicting that the computed ODE solution, 
at the given sample points, of course, is in fact within the requested accuracy and that, even
more surprisingly, the ODE solver LSODAR \emph{internally} proceeds adaptively.  In contrast, 
simply avoiding all these problems, LIMEX integrates fully adaptive, and Hermite interpolation 
of appropriate order is applied where necessary, strictly respecting the requested accuracy.  
Moreover, the fully adaptive approach (i.e.~its implementation in BioPARKIN, at least) seems 
to be much more efficient, see Table~\ref{Tab:01} and Figures~\ref{fig:_1},~\ref{fig:00}.

\subsection{GynCycle}

\paragraph{Description of the model.}
GynCycle is a differential equation model that describes the feedback mechanisms 
between a number of reproductive hormones and the development of follicles and 
corpus luteum during the female menstrual cycle \citep{pfizer}. 
The model correctly predicts hormonal changes following administration of 
single and multiple doses of two different drugs. 

\paragraph{BioPARKIN and the model.}
The model GynCycle is fairly large. It contains 33 species, 
2 assignment rules, 114 parameters, and 54 reactions.
The related benchmark timings for a forward simulation run and sensitivity 
calculations can be found in Table~\ref{Tab:01}.

Here, BioPARKIN served as a tool to explore the model and its parameter space. 
Together with its predecessor POEM (an unreleased, in-house tool based on the 
same numerical principles), it was able to develop and to fine-tune a highly 
descriptive and predictive model for a complex human pathway that has direct 
relevance to real-world applications.


\subsection{BovCycle}

\paragraph{Description of the model.}

The model BovCyle is a mechanistic mathematical model of the bovine estrous cycle 
that includes the processes of follicle and corpus luteum development and the key 
hormones that interact to control these processes \citep{Boer2011}.  The model 
generates a periodic solution without external stimuli, see Figure~\ref{fig:01}.  
The bovine estrous cycle is subject of extensive research in animal sciences.
Of particular interest have been, for example, the examination of follicular wave 
patterns \citep{Boer2011b}, as well as the study of synchronization protocols 
\citep{Claudia2011}.

\paragraph{BioPARKIN and the model.}

The BovCycle model consists of 15 species, 60 parameters, and 28 reactions.  
Again, the benchmark timings are given in Table~\ref{Tab:01}.  

In this application, BioPARKIN enabled the researchers to successively improve 
the model with each design iteration.  Procedures such as parameter identification 
and sensitivity analysis proved to be absolutely essential within this context as 
they guide design decisions by giving insight into hidden dependencies between parameters.

\subsection{BIOMD008} 

\paragraph{Description of the model.}

The model with ID 008 in the BioModels database describes the 
cell cycle control using a reversibly binding inhibitor.

\paragraph{BioPARKIN and the model.}

The model BIOMD008 comprises only 5 species, 21 parameters, and 13 reactions. 
The relevant benchmark timings for this model can also be found in Table~\ref{Tab:01}. 

Albeit being small, nevertheless, the model is of the cell cycle type and, 
in principle, exhibits a stable limit cyclic which is interesting by itself 
to look at sensitivity trajectories, see e.g.~Figure~\ref{fig:02}.

\paragraph{Parameter identification.}

Key questions of practical relevance in parameter identification tasks 
are almost always how much data is sufficient and, even more importantly, 
how much data is necessary to successfully identify the unknown parameters. 
We proceed as follows.

A specific parameter (V3p) is changed (from 0.3 to 1.0), and the goal is to 
reconstruct the original parameter value.  In a sequence of identification runs,
each of the five species is selected to be the only species for which data are 
available.  As data, we take the values of the selected species from the simulation 
run with the original parameter set, at the time points chosen adaptively by LIMEX. 

For three of the five species (M, Y, and Z), the original value of V3p is 
reconstructed without any difficulties.  The parameter identification, however, 
is not successful at all if one of the other two species (C and X) is selected 
as data source.

\paragraph{Sensitivities.}

We examined the sensitivity w.r.t.~parameter V3p.  The sensitivity overview 
for BIOMD008 results in a plot of the sensitivity trajectories of all species 
over time (see Figure~\ref{fig:02}).  Parameter V3p displays a cyclic sensitivity 
across all species.  It seems that a change in V3p influences the least the time 
course of species Y and Z while it has more influence on species C, M, and X.  
We note that these observations, apparently, are in distinct contrast to the 
findings of the parameter identification task just described.

\subsection{EpoRcptr}
\label{sub:EpoRcptr}

\paragraph{Description of the model.}

A dynamical model for the endocytosis of the erythropoietin receptor (EPO receptor) 
has been published in \cite{Hengl07}.  In fact, it is apparently a variant of BIOMD271 
of the database already mentioned above.  The model is relatively small as it consists 
of 7 species, 9 parameters, and 9 reactions.  However, there exist groups of functionally 
related parameters, that were identified by a statistical method in \cite{Hengl07}.  We 
use this example to demonstrate that BioPARKIN handles saddle points in the unknown 
parameter space correctly as opposed to, e.g., the Levenberg-Marquardt procedure that 
is well-known to not be able to detect these stationary points adequately.

\paragraph{BioPARKIN and the model.}

The model EpoRcptr is even smaller than BIOMD008, it contains 7 species, 9 parameters,
and 9 reactions.  The measurable values in this model, $Y_1, Y_2, Y_3$, are linear 
combinations of some species.  In BioPARKIN, these are added to the ODE system as 
algebraic equations, and thus forming a DAE system.  The integration routine LIMEX is 
capable of DAE systems up to order 1.  Again, the corresponding benchmark timings can 
be found in Table~\ref{Tab:01}.

\paragraph{Parameter identification.}

The parameter set as given in \cite{Hengl07} served as ,,true'' values of the model.
With these values the three measurement variables $Y_1, Y_2, Y_3$ have been sampled
by 10 equidistant points within the time interval $[0,100]$ each.  To be realistic,
5\% white, i.e.~normal distributed, noise has been added to this data set.   

For the identification run we took the time interval three times longer, 
$0 \leq t \leq 300$, and the true parameter values as initial guess for 
the iterative Gauss-Newton algorithm.  Since it is known that this point 
in parameter space lies on a lower dimensional manifold \citep{Hengl07}, 
the point has the character of a saddle point.  Indeed, identification 
runs of BioPARKIN indicate just this:  the higher $\mathrm{xtol}$ is chosen, 
the less iteration steps are made, reporting the stop at stationary points 
(i.e.~no reduction of the residual value) with unreasonably high incompatibility 
factors.  In addition, the initial parameter values (the ,,true'' values) are 
not recovered, but a different point on the parameter manifold is identified 
(Table~\ref{Tab:03}).  This can clearly be concluded by studying the related 
correlation matrix which contains in all cases a submatrix with entries near 
1 or -1 only.  In fact, the parameters $k_4$, $k_5$, and $k_6$ are thus 
connected by the correlation matrix, in total agreement with the findings 
as given in \cite{Hengl07}.

\paragraph{Sensitivities.}

The sensitivity trajectories of the measurement variables $Y_1$, $Y_2$, 
and $Y_3$ w.r.t.~parameters $k_4$, $k_5$, $k_6$ are depicted in 
Figures~\ref{fig:04},~\ref{fig:05}, and \ref{fig:06}, respectively.
As it can readily be seen, denser sampling of the measurement 
variables, especially for the variables $Y_1$ and $Y_3$, at later 
times should resolve the ambiguous parameter manifold.  Indeed, 
a convenient numerical test nicely confirms this conjecture, see 
Table~\ref{Tab:04}.

\begin{table} 
\caption{Parameter identification for model EpoRcptr.\label{Tab:03}}
\begin{indented}
\item[]
\begin{tabular}{@{}ccrccr}\br
  Parameter &  True Value &  Reconstruction  &      \multicolumn{3}{c}{Std.~Dev.}        \\\mr
     $k_1$  &    8.0e-03  &      8.114e-03   &  $\pm$ 2.053e-03  & $\hat{=}$ & 25.30 \%  \\
     $k_2$  &    5.0e-05  &      5.045e-05   &  $\pm$ 6.361e-06  & $\hat{=}$ & 12.61 \%  \\
     $k_3$  &    1.0e-01  &      1.012e-01   &  $\pm$ 8.970e-03  & $\hat{=}$ &  8.87 \%  \\
     $k_4$  &    2.5e-01  &      4.297e-01   &  $\pm$ 4.216e-03  & $\hat{=}$ &  0.98 \%  \\
     $k_5$  &    1.5e-01  &      1.096e-01   &  $\pm$ 2.732e-02  & $\hat{=}$ & 24.93 \%  \\
     $k_6$  &    7.5e-02  &      5.343e-02   &  $\pm$ 2.556e-02  & $\hat{=}$ & 47.83 \%  \\\br
\end{tabular}
\item[] Requested identification accuracy 
        has been $\mathrm{xtol} = 10^{-4}$.  Gauss-Newton iteration converged after 9 steps,
        with incompatibility factor $\kappa=0.04845$.
\end{indented}
\end{table}

\begin{table} 
\caption{Parameter identification for model EpoRcptr using more data.\label{Tab:04}}
\begin{indented}
\item[]
\begin{tabular}{@{}ccrccr}\br
  Parameter &  True Value &  Reconstruction  &      \multicolumn{3}{c}{Std.~Dev.}        \\ \mr
     $k_1$  &    8.0e-03  &      8.136e-03   &  $\pm$ 4.847e-04  & $\hat{=}$ &  5.96 \%  \\
     $k_2$  &    5.0e-05  &      4.956e-05   &  $\pm$ 1.702e-06  & $\hat{=}$ &  3.44 \%  \\
     $k_3$  &    1.0e-01  &      1.016e-01   &  $\pm$ 2.707e-03  & $\hat{=}$ &  2.67 \%  \\
     $k_4$  &    2.5e-01  &      2.546e-01   &  $\pm$ 1.215e-03  & $\hat{=}$ &  0.48 \%  \\
     $k_5$  &    1.5e-01  &      1.465e-01   &  $\pm$ 5.637e-03  & $\hat{=}$ &  3.85 \%  \\
     $k_6$  &    7.5e-02  &      7.201e-02   &  $\pm$ 2.443e-05  & $\hat{=}$ &  0.03 \%  \\ \br
\end{tabular}
\item[] Requested identification accuracy has been $\mathrm{xtol} = 10^{-4}$.
        Gauss-Newton iteration stopped at stationary point after 11 steps, 
        with incompatibility factor $\kappa=0.03227$.
              
\end{indented}
\end{table}

\subsection{A noteworthy caveat}

Key point, here, is that the sensitivity analysis is not always suitable to 
anticipate which parameters are more likely to be identified than others. 
In fact, sensitivities highly depend on the actual parameter set and, therefore, 
they are only meaningful at the end of a successful identification run. 
Thus, it really should always be kept in mind that the sensitivity results 
are merely meant as an exploratory a priori tool that might aid the researcher 
to get a better understanding of the model.

%
%

\section{Conclusion}

Systems biology as a scientific research field is getting more attention, 
and is gaining more practitioners around the world every year.  With the 
increased size of the community the importance of establishing standards 
becomes more pronounced.  The software package BioPARKIN presented here 
tries to inject long-standing mathematical experience into this growing 
community.  Ideally, this knowledge enables researchers to generate meaningful 
and reliable results even faster. 

While the computing time is comparable with other available software tools, 
BioPARKIN offers several unique features that are especially useful for 
biological modelling, such as breakpoint handling, or identifiability statements.
In particular, the implemented affine covariant Gauss-Newton method provides 
information on the compatibility between model and data, as well as on the 
uniqeness of a solution in case of convergence.  This is an important tool for 
model discrimination, when the ``best'' model is to be selected from several 
alternative models which all explain the given data equally well.  Moreover, 
the Jacobian can be computed with prescribed accuracy by solving the 
variational equation instead of using inaccurate numerical differentiation, 
thus increasing the reliability of numerical results.


\ack
This article is written in sincere remembrance of U.~Nowak who sadly 
passed away in June 2011.  Without his sophisticated contributions 
this work would have been clearly impossible.


\bibliographystyle{iopart-num}
\bibliography{arXiv_BioPARKIN}

\end{document}